%% file: 3rdqhrotationEuroPhys.tex
\documentstyle{europhys}
\input euromacr

\newcommand{\bm}[1]{\mbox{\boldmath$#1$}}
\begin{document}
%
%
%
\euro{}{}{}{}
\Date{}
\shorttitle{U. R. FISCHER \etal   
 HALL STATE QUANTIZATION IN A ROTATING FRAME}
%
%
%
\title{Hall state quantization in a rotating frame}
\author{Uwe R. Fischer and Nils Schopohl}
\institute{Eberhard-Karls-Universit\"at T\"ubingen, 
Institut f\"ur Theoretische Physik \\
Auf der Morgenstelle 14, D-72076 T\"ubingen, Germany }
%
%
\rec{}{in final form }
%
%
%
\pacs{
\Pacs{73}{40Hm}{Quantum Hall effect (integer and fractional)}
\Pacs{73}{50$-$h}{Electronic transport phenomena in thin films and low-dimensional structures}
      }
\maketitle
%
%
%
\begin{abstract}                
We derive electromagnetomotive force fields for charged particles moving 
in a rotating Hall sample, satisfying a twofold 
U(1) gauge invariance principle.   
It is then argued that the phase coherence property of 
quantization of the line integral of total collective particle momentum 
into multiples of Planck's quantum of action 
is solely responsible for quantization in the Hall state. 
As a consequence, the height 
of the Hall quantization steps should remain invariant 
in a rapidly rotating Hall probe.  
Quantum Hall particle conductivities do not depend on 
charge and mass of the electron, and are quantized 
in units of the inverse of Planck's action quantum.       
\end{abstract}




Modern molecular beam epitaxy enables the preparation of modulation-doped
semiconductor heterostructures in which, at low enough temperatures, a high
mobility two-dimensional electron gas is formed. 
This system is characterized by a long
Thouless dephasing length $l_{\phi }$,  
the distance within which phase coherence 
of mobile electrons is maintained. 
In good samples, and at low
enough temperatures, the length $l_{\phi }$ reaches several micrometers, 
exceeding the magnetic length $l_B=\sqrt{\hbar/eB}$ 
for applied magnetic fields of order one Tesla. 
Under these conditions, 
it should be possible to detect noninertial effects due 
to rotation or acceleration of the sample as a result of the change of 
quantum interference conditions, since 
the gauge potentials of the electromagnetic and noninertial 
fields experienced by the electrons both    
appear in 
their collective phase.
In what follows, we shall argue that quantum coherence  
under the influence of noninertial force fields is directly observable in 
the quantum Hall effect \cite{pers}. The quantum of Hall conductivity 
for particle transport is given by the inverse of 
Planck's quantum of action alone, and involves  
no properties specific to the electron. The  
arguments used to prove this result crucially rely
on the existence of a collective particle momentum expressing  
quantum coherence. By including gauge fields 
other than the electromagnetic one, we promote the idea that  
Hall quantization is of necessity 
derivable from a topological 
quantum number related to this coherence. It will be shown 
that this prediction about the nature of the quantum Hall effect 
is verifiable within current technological means.

The main quantity of interest to us 
is the total particle momentum 
\begin{equation}
{\bm p}= m{\bm v} + m {\bm \Omega}\times {\bm r}
+ q {\bm A}\,.
\end{equation}
Here, ${\bm A}$ 
is the electromagnetic vector potential, $q$ the 
charge of the electron {\it in vacuo}, ${\bm v}$  the particle velocity, 
and $m$ the inertial mass. 
The body is rigidly rotating with respect to the laboratory frame 
at a (time dependent) angular velocity $\bm \Omega$. 
The noninertial force on a massive test particle 
inside a rotating Hall probe, 
as measured in the rotating sample frame, is then 
given by the standard expression
\begin{equation}
{\bm F} = 
-2m{\bm \Omega} \times {\bm v} -m
{\bm \Omega}\times {\bm \Omega}\times {\bm r} 
- m\partial_t{\bm \Omega} \times {\bm r}. \label{noninertialforces} 
\end{equation} 
The first term on the right hand side represents 
(minus) the Coriolis force, 
the second one the centripetal force, and the last term is due to 
temporal changes of the angular velocity. The presence of this last term will 
prove to be crucial for our
argument on proper Hall quantization in a rotating frame 
presented below. We omit possible additional 
terms on the right hand side of 
equation (\ref{noninertialforces}) 
due to potential forces ({\it e.g.} gravity)   
or externally imposed linear acceleration.  
Vector and scalar potentials associated to rotation are defined as follows 
\begin{equation}
{\bm a}= {\bm \Omega} \times {\bm r}   \,,\qquad
a_0 = \frac12 \Omega^2 {\bm r}^2_\perp\,,  
\end{equation}
where ${\bm r}_\perp$ is the distance vector perpendicular to the axis of 
rotation. 
For a charged massive particle 
like the electron, we merge these potentials and 
the electromagnetic potentials into a generalized vector potential, 
incorporating the coupling 
constants charge $q$ and mass $m$,   
\begin{equation}\label{Adef} 
{\cal A} = q{\bm A} + m {\bm a}\,,
\end{equation}
and a generalized scalar potential 
\begin{equation}\label{chidef} 
{\chi} = -qA_0 - m a_0\,.
\end{equation}
The sum of the generalized {\em electromotive}
and {\em magnetomotive} forces, acting 
on an electron \cite{jackson}, 
consisting of noninertial plus proper Lorentz and 
electric forces, then takes on the form 
\begin{eqnarray}
{\bm F}_{{\cal L}} = 
{\cal E} +{\bm v}\times {\cal B}\,,
\label{genLorentz}
\end{eqnarray}
where the generalized electric and magnetic fields are
\begin{eqnarray}
{\cal E} &=& -\nabla\chi -\partial_t {\cal A}\,,\nonumber\\
{\cal B} &=& \nabla \times {\cal A}\,.\label{genFields}
\end{eqnarray}
As a consequence of this relation for the total force, 
the usual expression for the drift velocity of the charge carriers,   
resulting from zero total force  
in perpendicular electric and magnetic fields,  
experiences the obvious modification that ${\bm E}\rightarrow {\cal E}$ 
and ${\bm B}\rightarrow {\cal B}$, so that  
${\bm v}_D= {\cal E}\times {\cal B}/{\cal B}^2$.

The 
force fields displayed in equations (\ref{genLorentz}) and
(\ref{genFields}) 
give a theory possessing in effect two U(1) gauge symmetries. 
The standard U(1) from electromagnetism, with coupling
constant $q$ (charge), 
and another U(1) gauge symmetry, with coupling constant 
$m$ (inertial rest mass).  
The gauge 
potential of this second U(1) has 
a scalar part $a_0$ and a vectorial part ${\bm a}$. 
The homogeneous Maxwell equations 
${\rm rot}\, {\cal E}= -\partial_t {\cal B}$ and 
${\rm div}\, {\cal B} = 0$ then follow from the existence of the potentials
$\cal A$ and $\chi$ in (\ref{Adef}) and (\ref{chidef}).  That the Faraday
law holds is due to our admitting a variation of the angular velocity with 
time and the resulting last force term in (\ref{noninertialforces}). 
This important term, leading 
to gauge invariance in explicitly time dependent situations,
 has not been considered in \cite{johnsonAJP}, where 
the Hall effect under rotation, without the simultaneous 
existence of a magnetic field, was investigated.   


The gauge invariant particle current 
induced by the electromotive force 
field $\cal E$ is in linear response ($a,b \in\{x,y\}$): 
\begin{eqnarray}
{\bm J}_a^{\rm ind} & = & {\bm {\tilde \sigma}}_{ab}{\cal E }_b
\nonumber\\
& = & {\bm {\tilde \sigma}}_{ab} \left( q{\bm E} + m{\bm g} \right)_b 
\,, 
\label{sigmadef}
\end{eqnarray}
where ${\bm g}=\nabla a_0-\partial_t {\bm a}\,$
represents the `electric' part of the mechanical acceleration 
experienced by the electron. In the present case this acceleration 
is purely caused by rotation, and takes the form 
${\bm g} = -{\bm \Omega}\times {\bm \Omega}\times {\bm r} 
-\partial_t{\bm \Omega}\times {\bm r}$. 

Observe that the left hand side of equation (\ref{sigmadef}) contains the  
number current density rather than the electric current density 
and that, dimensionally, the particle conductivity 
$[\tilde \sigma ] = [\sigma_{el}/q^2]$. 
In the case of two coupling constants, $m$ and $q$, it is the 
number of particles crossing (in two spatial dimensions) 
a line of unit length per unit time, which is 
the relevant observable. This quantity is proportional to the 
electromotive force field $\cal E$, 
which causes these particles to move. 
If we were to use the transport
coefficient $\sigma_{\rm el}$ and the transport equation 
${\bm J}_{\rm el}^{\rm ind} = {\bm {\sigma}}_{\rm el}{\bm E }$, 
a rotating Hall sample does not yield the sharp 
conductivity quantization steps observed in nonrotating samples. 
We will show below that Hall state quantization is, according to equation 
(\ref{sigmaquant}), to be expressed in the particle 
Hall conductivity $\tilde \sigma_{xy}$ occuring in relation 
(\ref{sigmadef}).  

Evidence for the necessity of 
using the particle transport equation (\ref{sigmadef}) 
comes from the existence of the 
London field in superconductors.  
Complete expulsion of the field $\cal B$ deep inside in a superconductor 
requires the particle conductivity $\bm {\tilde \sigma}$ 
to have a contribution proportional to 
$1/i\omega$, which yields a term on the right hand side
of (\ref{sigmadef}), proportional to the generalized vector potential 
$\cal A$. 
Corresponding to complete Meissner type screening,  
${\cal B}={\rm rot}\, {\cal A}= q {\bm B} + 2m{\bm \Omega}={0}$, 
the London spontaneous field ${\bm B}_L$ 
then takes the value 
\begin{equation}\label{London}
{\bm B}_L = -2\frac{m} q {\bm \Omega}\,.
\end{equation}
This 
relation corresponds to zero winding number of the 
phase $\theta$, cf. equations (\ref{pdqInt})--(\ref{PhiEq}) below. 
Equation  (\ref{London})  
has been verified experimentally already 35 years ago 
\cite{zimmerman}, in an experiment  
in which it was used to infer 
the Compton wavelength of superconducting electrons. 
For the linear in velocity (nonrelativistic) 
limit and in a superconductor, 
the Cooper pair mass $m$  
equals twice the electron inertial 
rest mass {\it in vacuo}, $2m_e$ 
(the outcome of a more 
recent high precision experiment using a rotating superconducting niobium ring
\cite{tatecabrera} has been $m/2m_e=1.000084(21)$).   
If we insert on the left hand side of the equation (\ref{London}) 
the bare electron values
$m=2m_e$ and $q = -2e$ ($e>0$), 
we have 
${\bm B}_L = (1.14 \cdot 10^{-11} 
{\rm Tesla} \cdot {\rm sec}) \, {\bm \Omega} \,$. 
Only the ratio of $m$ and $q$ enters the London induced 
magnetic flux strength. In a quantum Hall liquid, where the ``elementary'' 
quanta are $q=-e$, $m=m_e$ and $\phi_0=2\pi\hbar/e$, 
instead of $q=-2e$, $m=2m_e$ and $\phi_0=2\pi\hbar/2e$ in 
a superconductor, the London flux strength thus takes for a given 
$\bm \Omega$ the same value (\ref{London}) like 
in the superconductor.
That the mass is exactly the bare inertial 
mass to linear order in the particle velocity is independently 
substantiated by a recent discussion of the London equation 
in \cite{londoneqliu}, by using (thermodynamic) arguments different  
from our gauge invariance argument.  
If the quantum Hall fluid is described in a (relativistic)
theory with interactions, 
the particle mass is to be replaced by the chemical potential, cf. 
\cite{annalspaper}, but the identity of $m$ with the bare mass 
to linear order in the velocity still persists. 

Quantum coherence properties enter the stage 
if we require for the line integral 
of collective particle momentum along a closed path  
\begin{equation}\label{pdqInt}
\oint {\bm p}\cdot d{\bm r} = N_v 2\pi \hbar \,,  
\end{equation}
where $N_v$ is the winding number of 
phase $\theta$, 
such that the total canonical 
momentum 
\begin{eqnarray}
{\bm p} & \equiv &\hbar \nabla \theta
\nonumber\\
& = & m{\bm v} + m {\bm \Omega}\times {\bm r}
+ q {\bm A} \nonumber\\
& = & m{\bm v}  + {\cal A}\,.
\end{eqnarray}  
The uniqueness condition of the collective 
phase 
represented in (\ref{pdqInt}) then leads, if we
take a path in the bulk of the 
electron liquid, for which the integral of  
$m{\bm v}$ may be neglected, to the quantization of the sum of a 
Sagnac flux \cite{PostSagnac,stedman}  
and the magnetic flux:
\begin{eqnarray} 
\Phi & = & q\oint {\bm A}\cdot d{\bm r} + m \oint {\bm \Omega}\times {\bm r}
\cdot d{\bm r}
\nonumber\\
& = & \int\!\!\int {\cal B}\cdot d{\bm S}= N_v 2\pi \hbar \,.  \label{PhiEq} 
\end{eqnarray} 
This flux quantization rule associated with the field $\cal B$  
corresponds to the fact that a {\em vortex} 
is fundamentally characterized by the winding number $N_v$ 
alone \cite{annalspaper}.  
No properties of the medium in which it lives, in particular the mass 
and charge of the medium's constituents, enter the quantum of generalized 
flux. 

Consider now the (purely magnetic) 
filling factor of a nonrotating two-dimensional electronic 
system of areal density $n_{2D}$ 
in a large magnetic field $B$ at low temperatures.  
The ratio $\nu = {n_{2D}}/({B /\phi_0})$, where $\phi_0=2\pi \hbar/e$,
gives the inverse of the number of singly quantized ($N_v=1$) magnetic
flux quanta available per electron.  
The Hall resistance is quantized into $R_H=(2\pi \hbar/e^2) \nu^{-1}
=R_K /\nu$ (the von Klitzing constant $R_K=25812.807$ $\Omega$), 
with integer 
or fractional $\nu$ 
\cite{dctsuiQH}. 
We have seen that the dynamical, generalized 
magnetic force field occuring in the Hamiltonian in the noninertial 
rotating state is $\cal B$. 
The solution of the Landau problem for the electronic energy levels 
in a magnetic field thus refers to a Hamiltonian $H= ({\bm p} -{\cal A})^2/2m
+ \chi $, depending on the generalized vector potential $\cal A$ and 
magnetic field $\cal B$.
Hence, in the noninertial case, 
the Landau level degeneracy per unit area is ${\cal B} /(2\pi \hbar)$,   
and contains the generalized magnetomotive field $\cal B$ instead of 
the magnetic flux strength ${\bm B}$. By assigning this  
value to the degeneracy, use is made of the fact that 
the ``magnetic length'' associated with rotation, 
$l_\Omega \equiv \sqrt{\hbar/(2m\Omega)}$, 
is much larger than the Thouless length as well as 
the proper magnetic length, $l_\Omega\gg l_\phi > l_B$ \cite{RemarkLength}. 
As a consequence, the linear 
dependence on position of the part of 
$\cal E$ associated with rotation, 
given by $-m{\bm \Omega}\times {\bm \Omega}\times {\bm r} 
- m\partial_t{\bm \Omega} \times {\bm r}$, does not lift the Landau
level degeneracy beyond the broadening of the levels 
already taking place due to scattering.
Under the condition $l_\Omega\gg l_\phi > l_B$, the filling factors  
assigned to the Landau levels are    
\begin{equation}
\nu_{\cal B}=\frac{n_{2D}}{{\cal B}/(2\pi \hbar)}\,.
\end{equation}
The Faraday law ${\rm rot}\, {\cal E} = -\partial_t {\cal B}$, telling
us how the flux strength 
corresponding to the vector potential $\cal A$ changes in 
time, is in its integrated form 
\begin{equation}\label{homMaxwell}
\oint_\Gamma {\cal E}\cdot d{\bm r} = - \frac{d\Phi}{dt}\,.   
\end{equation}  
Consider the adiabatic process of slowly turning on a flux quantum  
$2\pi \hbar$ in a nondissipative Hall 
state, which has $\tilde\sigma_{xx}=\tilde\sigma_{yy}=0=
\tilde\rho_{xx}=\tilde\rho_{yy}$, and the antisymmetry property  
$\tilde\sigma_{xy}=-\tilde\sigma_{yx}$ \cite{pers}. 
The path $\Gamma$ is led around the flux tube.
The induced current then obeys ${\hat z}\times {\bm J}^{\rm ind} 
= \tilde\sigma_{xy} {\cal E}$, and 
the number of particles $N$ inside the area enclosed by 
$\Gamma$ changes according to 
\begin{equation}
\frac{dN}{dt} = \tilde\sigma_{xy} \frac{d\Phi}{dt}\,.
\end{equation} 
After a single quantum of  generalized flux $2\pi \hbar$ 
has been added, and because 
the number of particles is integral, this implies 
that the off-diagonal part of the particle transport 
conductivity defined in (\ref{sigmadef}) is quantized according to 
\begin{equation}\label{sigmaquant}
\tilde\sigma_{xy}= \nu_{\cal B} /2\pi \hbar\,, 
\end{equation}
with $\nu_{\cal B}$ an integer \cite{fractional}.
Hence, to summarize this derivation, 
the generalized Faraday law in a rotating frame
gives in a nondissipative Hall state 
the quantization of the Hall conductivity if and only if the 
moving generalized flux in (\ref{PhiEq}) is quantized in units of the 
action quantum. This argument is similar to the one given by Laughlin 
\cite{laughlin1}, extended to a rotating frame.


The 
quantization of the Hall resistance 
into $R_K/\nu$ for nonrotating samples has been measured to 
an absolute accuracy of a few parts in 
$10^{-8}$ for an individual, specific sample, whereas 
in a comparative study of different materials, a relative accuracy of about 
$10^{-10}$ of the ratio of Hall resistances has been achieved 
\cite{compareQH}. Considering that magnetic fields in quantum Hall 
experiments cover a range $B\sim 1\cdots 30$ Tesla,  
this implies that with a rotation rate of the Hall sample of, say, 
$\Omega = 10^3\, {\rm sec}^{-1}\cdots 10^4\, {\rm sec}^{-1}$, the Sagnac 
contribution in (\ref{PhiEq})  
is large enough to verify if Hall quantization experiences
a change if the sample rotates. 
For the comparative measurement, the rotation rates required are lower 
by about two orders of magnitude. 
At the very low temperatures of order $10^{-3}$ K needed for superfluid 
$^3\!$He, rotation rates of the cryostat of 
order $\Omega\sim 10 $ sec$^{-1}$ %
already have been realized \cite{mattireview}.
In relation to superfluid $^3\!$He, it is also worthwhile to mention here  
that for thin $^3\!$He-A films, an electrically
neutral system, a half integer quantum Hall effect, owing to a topological 
invariant of the $p$-wave order parameter in this system, has been suggested 
\cite{grishaQH3He}. 

We point out that with respect to the interpretation of  
experimental results, it should be borne in mind 
that the force fields in (\ref{genLorentz}) and (\ref{genFields}) 
refer to the rotating sample frame, and not to the laboratory frame.    

The experiment proposed, then, consists in a comparison 
of the quantum Hall resistances 
in the reference frame of a rotating sample 
as well as in a nonrotating sample. 
If identical quantization results are obtained, 
this yields direct proof that what is actually observed in the Hall experiment
is the quantization according to (\ref{sigmaquant}), 
rather than quantization into $e^2/2\pi\hbar$. 
The electric charge of the electron, 
the U(1) coupling constant of electromagnetism, enters 
if we count the number of particles arriving at the Hall contacts,  
by ascribing to their transport 
properties an electric conductance in an electric circuit. What is invariantly 
measured, though, is the induced number current in (\ref{sigmadef}). 
Particle number currents can be distinguished from conventional electric 
currents as follows. Whereas electrical currents are measured by (classical) 
means of an impedance 
and the ensuing voltage drop, intrinsic particle currents 
can be measured if we use tunneling contacts, for which the 
wave-particle duality embodied in the tunneling process 
ensures that particles are counted. 

We stress that if we were to use  
a Chern-Simons effective field theory for a description of 
the quantum Hall effect \cite{wilczekanyons},  such an effective
description is expressible within our approach.
For that purpose, one uses a term in the action density of the form
\begin{equation}
-({J}^{\rm ind})^\mu {\cal A}_\mu + 
\frac14 \tilde \sigma_{xy}\epsilon^{\alpha\beta\gamma} 
{\cal A}_\alpha {\cal F}_{\beta\gamma}\,,
\end{equation} 
where $\epsilon^{\alpha\beta\gamma}=\pm 1 $ (with the sign convention 
$\epsilon^{0xy}=+1$) is the unit antisymmetric symbol in three space-time 
dimensions and 
${\cal F}_{\mu\nu}=\partial_\mu {\cal A}_\nu -\partial_\nu {\cal A}_\mu$ 
is the field tensor constructed from $\cal E$ and $\cal B$. 
Such a term reproduces the induced current (\ref{sigmadef}) 
in the nondissipative Hall state as a result of varying the action 
with respect to ${\cal A}_i$.  
Integration of the zeroth component of the complete field equations 
$
({J}^{\rm ind})^\mu = \frac12  \tilde \sigma_{xy}
\epsilon^{\mu\alpha\beta} {\cal F}_{\alpha\beta}
$, giving the induced density 
\begin{equation}
\rho^{\rm ind} = \tilde \sigma_{xy}{\cal B}_z\,,
\end{equation}
tells us that each particle associated with 
$\rho^{\rm ind}$
carries generalized flux 
$1/\tilde \sigma_{xy}=2\pi \hbar /\nu_{\cal B}$. 

In conclusion, the quantization of the Hall particle 
conductivity under rotation has been derived 
by invoking the following basic arguments. 
(i) The generalized 
Lorentz force equation (\ref{genLorentz}), containing the invariant
field strengths $\cal E$ and $\cal B$, and  
 describing the motion of the charge carriers,  is valid.
(ii) The flux conservation law for the magnetomotive field
$\cal B$ in  (\ref{homMaxwell}) holds true.
(iii) We consider a nondissipative Hall state, which has
$\tilde\sigma_{xx}=\tilde\sigma_{yy}=0=\tilde\rho_{xx}=\tilde\rho_{yy}$.
(iv) A necessary condition for Hall quantization to hold is that 
the total collective canonical particle momentum is derivable from
a collective quantum phase, ${\bm p}\equiv \hbar \nabla \theta$, 
such that the Bohr-Sommerfeld type integral of this momentum 
is quantized into units of Planck's action quantum.
The translation of the conventional quantum Hall problem 
into a rotating frame elucidates and strengthens the point of view 
that it is essentially a basic 
Hamiltonian quantity in phase space, 
the closed action integral of the collective  
particle momentum $\bm p$, 
which yields Hall state quantization.

An experimental proof or disproof of the quantization rule 
(\ref{sigmaquant}) in a rotating system will show if our assertion about the 
nature of the quantum Hall phenomenon is true and that it indeed 
relies upon the existence 
of collective particle momentum and the motion of the 
associated generalized flux quanta.

We thank Grisha Volovik 
for a discussion of 
the contents of this work. 
U. R. Fischer 
acknowledges financial support 
by the DFG (FI 690/1-1).

\end{document}

%% file: euromacr.tex
\def\etal{{\hbox{{\tenit\ et al.\/}\tenrm :\ }}}

\newif\ifboo \boofalse
